\documentclass[11pt]{article}
\usepackage{hyperref}
\usepackage{amsmath}
\usepackage{graphicx,epsfig}
\usepackage{subfigure}
\usepackage{color}
\usepackage{bm}
\usepackage{multirow}
\usepackage{verbatim}
\usepackage{comment}
\usepackage{url}
\usepackage{amssymb}
\usepackage{amsfonts}
\usepackage{amsmath}

\usepackage{authblk}

\usepackage{enumitem}

\usepackage[displaymath]{lineno}

\usepackage{multirow}

\usepackage{graphics}
\usepackage{epsfig}
\usepackage{pgf}
\usepackage{pgfpages}
\usepackage{xcolor}
\usepackage{pifont}

\graphicspath{{./Figs/}}

\usepackage{float}     % locating the picture exactly at that place

\usepackage{titletoc}
\usepackage[raggedright,pagestyles]{titlesec}

\usepackage{fancyhdr}
\usepackage{lastpage}
\usepackage{layout}
\usepackage[footnotesep=10mm]{geometry}
\usepackage{caption2}

\usepackage{setspace}

\makeatletter
\renewcommand\@biblabel[1]{\footnotesize$[{#1}]$\selectfont}
\makeatother  %

\titlecontents{part}[0em]{\vskip3mm \bf B.\contentslabel{9em}}{}
{\bf \hspace*{0em}}
{\titlerule*[0.8pc]{.}\contentspage}[\addvspace{0.5ex}]

\titlecontents{section}[1em]{\it \contentslabel{0em}}{\hspace*{1.5em}}
{\it\hspace*{2.3em}}
{\titlerule*[0.8pc]{.}\contentspage}[\addvspace{0.5ex}]

\titlecontents{subsection}[2.7em]{\contentslabel{0em}}{\hspace*{2.5em}}{\hspace*{2.3em}}
{\titlerule*[0.8pc]{.}\contentspage}[\addvspace{0.5ex}]

\titleformat{\part}{\raggedright\large\bfseries}{\bf\normalsize{B.\thepart} }{1em}{}
\titleformat{\section}{\raggedright\normalsize\bfseries}{\bf\normalsize{\S\,\thesection}.}{1em}{}
\titleformat{\subsection}{\raggedright\normalsize\bfseries}{\thesubsection}{1.5em}{}
\titleformat{\subsubsection}{\raggedright\normalsize\itshape}{\thesubsubsection}{1em}{}

\usepackage{hyperref}

\linethickness{0.6pt}

\newcommand\PICorientline{\setlength{\unitlength}{2.5mm}
\begin{picture}(2,1)
\put(0.9,-0.5){\vector(0,1){2}}
\end{picture}}

\newcommand\PICorientcircleright{\setlength{\unitlength}{3mm}
\begin{picture}(1.8,1)(0.2,0.5) \put(1,1){\circle{1.2}}

 \put(1.3,1.55){\vector(1,0){0.01}}   %%% Put arrow to the circle

\end{picture}}

\newcommand\PICorientpluscross{\setlength{\unitlength}{2mm}
\begin{picture}(2.9,2)(-1.4,-0.6)
\put(-1,-1){\vector(1,1){2.3}}%
\put(-1,1){\vector(-1,1){0.3}}%
\qbezier(-0.3,0.3)(-0.8,0.8)(-1.0,1.0)
\qbezier(0.3,-0.3)(0.8,-0.8)(1.0,-1.0)
\end{picture}}

\newcommand\PICorientminuscrossrotated{\setlength{\unitlength}{2mm}
\begin{picture}(2.9,2)(-1.4,-0.6)
\put(-1,-1){\vector(1,1){2.3}}%
\put(1,-1){\vector(1,-1){0.3}}%
\qbezier(-0.3,0.3)(-0.8,0.8)(-1.0,1.0)
\qbezier(0.3,-0.3)(0.8,-0.8)(1.0,-1.0)
\end{picture}}

\newcommand\PICorientminuscross{\setlength{\unitlength}{2mm}
\begin{picture}(2.9,2)(-1.4,-0.6)
\put(1,-1){\vector(-1,1){2.3}}%
\put(1,1){\vector(1,1){0.3}}%
\qbezier(0.3,0.3)(0.8,0.8)(1.0,1.0)
\qbezier(-0.3,-0.3)(-0.8,-0.8)(-1.0,-1.0)
\end{picture}}

\newcommand\PICorientLRsplit{\setlength{\unitlength}{0.5mm}
\begin{picture}(10,8)(-1.5,2)
\qbezier(0,0)(4,4)(0,8)
\qbezier(6,0)(2,4)(6,8)
\put(-0.2,9){\vector(-2,3){0.2}}%
\put(7,9){\vector(2,3){0.2}}
\end{picture}}

\newcommand\PICorientUDsplit{\setlength{\unitlength}{0.6mm}
\begin{picture}(11,6)(-1,1)
\qbezier(0,6)(4,2)(8,6) \qbezier(0,0)(4,4)(8,0)
\put(9,6.6){\vector(3,2){0.2}}%
\put(9,-0.4){\vector(3,-2){0.2}}
\end{picture}}

\newcommand\PICorientopengammaplus{\setlength{\unitlength}{0.5mm}
\begin{picture}(12,8)(-2.5,-6)
\qbezier(1.4,-5.2)(2.0,-6.2)(3.2,-6.1)
\qbezier(3.2,-6.2)(6.0,-6.2)(6.2,-3.1)
\qbezier(6.2,-3.1)(5.9,-0.3)(3.1,-0)
\qbezier(3.1,-0)(0.3,-0.3)(0.0,-2.9)
\put(0.0,-3.1){\line(0,-1){7}} \put(0,0.8){\vector(0,1){4}}
\end{picture}}

\newcommand\PICorientopengammaminus{\setlength{\unitlength}{0.5mm}
\begin{picture}(12,8)(-2.5,0)
\qbezier(1.4,5.2)(2.0,6.2)(3.2,6.1)
\qbezier(3.2,6.2)(6.0,6.2)(6.2,3.1)
\qbezier(6.2,3.1)(5.9,0.3)(3.1,0)
\qbezier(3.1,0)(0.3,0.3)(0.0,2.9)
\put(0.0,3.1){\vector(0,1){7}} \put(0,-0.8){\line(0,-1){4}}
\end{picture}}

\newcommand\PICcircle{\setlength{\unitlength}{3mm}
\begin{picture}(1.8,1)(0.2,0.5) \put(1,1){\circle{1.2}}
\end{picture}}

\newcommand\PICunorientLRsplit{\setlength{\unitlength}{0.6mm}
\begin{picture}(9,6)(-1,2)
\qbezier(0,0)(4,4)(0,8) \qbezier(6,0)(2,4)(6,8)
\end{picture}}

\newcommand\PICunorientUDsplit{\setlength{\unitlength}{0.6mm}
\begin{picture}(11,6)(-1,1)
\qbezier(0,6)(4,2)(8,6) \qbezier(0,0)(4,4)(8,0)
\end{picture}}

\newcommand\PICunorientpluscross{\setlength{\unitlength}{2mm}
\begin{picture}(2.9,1.9)(-1.3,-0.6)
\qbezier(-1,-1)(0.0,0.0)(1,1)
\qbezier(-0.3,0.3)(-0.8,0.8)(-1.0,1.0)
\qbezier(0.3,-0.3)(0.8,-0.8)(1.0,-1.0)
\end{picture}}

\newcommand\PICunorientminuscross{\setlength{\unitlength}{2mm}
\begin{picture}(2.9,1.9)(-1.3,-0.6)
\qbezier(1,-1)(0.0,0.0)(-1,1)
\qbezier(0.3,0.3)(0.8,0.8)(1.0,1.0)
\qbezier(-0.3,-0.3)(-0.8,-0.8)(-1.0,-1.0)
\end{picture}}

\newcommand\PICclosegammaplus{\setlength{\unitlength}{0.5mm}
\begin{picture}(16,8)(-8,-2)
\qbezier(1.4,-1.8)(2.0,-2.7)(3.2,-2.9)
\qbezier(3.2,-2.9)(6.0,-2.7)(6.2,-0.2)
\qbezier(6.2,-0.2)(5.9,2.6)(3.1,2.9)
\qbezier(3.1,2.9)(0.3,2.6)(0.0,0.0)
\qbezier(-1.4,1.8)(-2.0,2.7)(-3.2,2.9)
\qbezier(-3.2,2.9)(-6.0,2.7)(-6.2,0.2)
\qbezier(-6.2,0.2)(-5.9,-2.6)(-3.1,-2.9)
\qbezier(-3.1,-2.9)(-0.3,-2.6)(0.0,0.0)
\end{picture}}

\newcommand\PICclosegammaminus{\setlength{\unitlength}{0.5mm}
\begin{picture}(16,8)(-8,-2)
\qbezier(-1.4,-1.8)(-2.0,-2.7)(-3.2,-2.9)
\qbezier(-3.2,-2.9)(-6.0,-2.7)(-6.2,-0.2)
\qbezier(-6.2,-0.2)(-5.9,2.6)(-3.1,2.9)
\qbezier(-3.1,2.9)(-0.3,2.6)(0.0,0.0)
\qbezier(1.4,1.8)(2.0,2.7)(3.2,2.9)
\qbezier(3.2,2.9)(6.0,2.7)(6.2,0.2)
\qbezier(6.2,0.2)(5.9,-2.6)(3.1,-2.9)
\qbezier(3.1,-2.9)(0.3,-2.6)(0.0,0.0)
\end{picture}}

\newcommand\PICorientUDinconsistentsplit{\setlength{\unitlength}{0.6mm}
\begin{picture}(11,6)(-1,1)
\qbezier(0,6)(4,2)(8,6)
\qbezier(0,0)(4,4)(8,0)
\put(-1,7){\vector(-4,3){0.2}}%
\put(9,7){\vector(4,3){0.2}}
\put(-1,-1){\vector(-4,-3){0.2}}%
\put(9,-1){\vector(4,-3){0.2}}
\end{picture}}

\begin{document}
\title{Topological invariants for superconducting cosmic strings}
\author[1]{\small Xinfei LI}
\author[2,1]{\small Xin LIU \thanks{Corresponding author: xin.liu@bjut.edu.cn}}
\affil[1]{\small Institute of Theoretical Physics, Beijing University of Technology,  Beijing 100124, China}
\affil[2]{\small Beijing-Dublin International College, Beijing University of Technology,  Beijing 100124, China}
\date{\today}
\maketitle
\begin{abstract}
Superconducting cosmic strings (SCSs) have received revived interests recently. In this paper we treat closed SCSs as oriented knotted line defects, and concentrate on their topology by studying the Hopf topological invariant. This invariant is an Abelian Chern-Simons action, from which the HOMFLYPT knot polynomial can be constructed. It is shown that the two independent parameters of the polynomial correspond to the writhe and twist contributions, separately. This new method is topologically stronger than the traditional (self-) linking number method which fails to detect essential topology of knots sometimes, shedding new light upon the study of physical intercommunications of superconducting cosmic strings as a complex system.
\end{abstract}

%%%%%%%%%%
\section{Introduction}

Cosmic strings are thought to be created as one-dimensional topological defects from symmetry breaking phase transition in the early universe from the field theoretical point of view \cite{Kibble:1976}. String theory provides them another formation machinery that they may arise from brane inflation scenarios \cite{Sarangi:2002yt} (such as D- and F-strings \cite{Copeland:2003bj}) which are able to avoid two main unpleasant prospects, \textit{ high string tension} and \textit{dilution after inflation}. See \cite{Kibble:2013} for a review.

In 1985 Witten proposed a model of superconducting cosmic strings (SCSs) \cite{Witten:1985}. It contains two complex scalar fields and two vector fields, producing cosmic strings under a symmetry-breaking mechanism $U(1)\otimes U(1) \to U(1)$. Massless charges  condensate and distribute in the vicinity of string core, thus cosmic strings behave in a superconducting wire manner.
%so does the D-brane \cite{Dvali:2003zj}.
These strings' motion is expected to give rise to bosonic or fermionic currents within cosmic magnetic plasma or a primordial magnet field, leading to a theory that SCS radiation may cause cosmic gamma burst rays \cite{Vilenkin:1994}. In recent years SCSs have received renewed interests. Researches suggest that radiation of
SCS cusps could be responsible for the gamma burst \cite{Vachaspati:2008su,Yu:2014gea,Gruzinov:2016hqs}, and SCSs are able to arouse CMB spectral distortion \cite{Tashiro:2012nb} although they have only a small percentage in thermal fluctuation \cite{Lazanu:2014eya,Hindmarsh:2011}. Topology of SCSs is closely related to their energy and other dynamical properties, especially in the study of physical intercommunications of strings. A mechanism is discovered that the collapse of a knotted structure (texture) distributed in a spherical region produces a metric perturbation, which could lead to hot and cold spots in cosmic microwave background \cite{Turok:1990,Turok:2007}.

In this paper we focus on the topology of SCSs, by regarding them as oriented knots and studying their Hopf mapping degree.
The Hopf degree is a topological invariant of the abelian Chern-Simons type, based on which the HOMFLYPT polynomial knot invariant can be successfully constructed \cite{HOMFLY:1985,PT:1987}. The HOMFLYPT is a two-parameter topological invariant more powerful than the tool of traditional (self-)linking numbers which sometimes fail to detect essential topology of knots. Hopefully our new method will find important applications in classification of complex tangles of SCSs as well as physical reconnection processes during evolutions of the strings.

The paper is arranged as follows. In Section 2, the SCS model is presented which generates bosonic currents and one-dimensional string configurations. In Section 3, the Hopf mapping degree as a Chern-Simons type action is introduced for the purpose of investigating the knotted string structures. We will show that the traditional (self-)linking numbers stemming from this Hopf invariant are weak in detecting topology of oriented knotted comic strings. In Section 4, we will derive from the Hopf invariant the skein relations of the Kauffman R-knot polynomial, showing that the two independent parameters of the R-polynomial originate from the individual contributions of writhe and twist of the strings. Then, the two-variable HOMFLYPT knot polynomial can be subsequently constructed from the R-polynomial. In Section 5, some typical examples are provided for reader's convenience, and conclusion and discussion are finally presented in Section 6.

\section{Cosmic superconducting strings}
SCSs models carry bosonic or fermionic currents. In this paper we consider only the bosonic currents produced by superconducting strings;  the study of fermionic currents is similar \cite{Witten:1985,Vilenkin:1994}. The action reads
\begin{equation}
  S=\int dx^4 \left[
  \left|\widetilde D_{\mu}\phi\right|^2+\left| D_{\mu}\sigma\right|^2-\frac{1}{4}\widetilde F^{\mu\nu} \widetilde F_{\mu\nu}-\frac{1}{4} F^{\mu\nu} F_{\mu\nu}+ V\left(\phi,\sigma \right) \right],
 \quad \mu,\nu = 1,2,3,4,      \label{1stActn}
  \end{equation}
where
  \begin{equation}
   V\left(\phi,\sigma \right)= \frac{1}{4}\lambda_{\phi}\left(\left|\phi\right|^2-\eta^2_{\phi}\right)^2    +\frac{1}{4}\lambda_{\sigma}\left(\left|\sigma\right|^2-\eta^2_{\sigma}\right)^2+\beta \left| \phi \right|^2\left| \sigma \right|^2.
\end{equation}
This is a double complex scalar model with $U(1) \otimes U(1)$ symmetry, $\phi$ and $\sigma$ being the two complex scalar fields. Their covariant derivatives are, respectively, $D_\mu=\partial_\mu -ig A_\mu$ with coupling constant $g$ and $U(1)$ gauge potential $A_\mu$, and $\widetilde D_\mu=\partial_\mu -ie\chi_\mu$ with coupling $e$ and potential $\chi_\mu$. The pairs $\left(\eta_\phi,\lambda_\phi\right)$ and $\left(\eta_\sigma,\lambda_\sigma\right)$ indicate the zeroes of the $\phi$ and $\sigma$ fields, respectively, and $\beta$ represents their mutual coupling. This model is able to provide a source for dark energy quintom \cite{Elizalde:2004mq,Cai:2009zp}.

Let the $\phi$-related $U(1)$ symmetry break in the ground state, and the $\sigma$-related symmetry remain to serve as an electromagnetic field. The charge condensation on the cores of cosmic strings depends on the fine-tuning of the Higgs sector; the current in a core is less than the critical level to keep the stability of the current-carrying of strings. SCSs are oriented strings with the electromagnetical currents most confined in them \cite{Witten:1985,Vilenkin:1994}. Locally, an SCS acts as a cylinder of width $m^{-1}_\phi$ formed by the $\phi$ symmetry-breaking; globally, it is very thin when observed from far away. In this paper we study SCSs as oriented topological line defects.

The action \eqref{1stActn} leads to an equation of motion (EOM) for $\phi$.
Given a boundary condition, the $\phi$ field can be solved out whose distribution indicates the existence of topological defects on the manifold. \textit{See a note on this treatment in the end of this section.}

To extract the information of the topological defects from the $\phi$ distribution,  let $\phi$ \textit{induce} a covariant derivative $D_{\mu}^{\text{ind}}$ through a parallel field condition,
\begin{equation}
D^{\text{ind}}_{\mu}\phi=\partial_{\mu}\phi - i\alpha_A A^{\text{ind}}_\mu\phi = 0.     \label{ParallCond}
\end{equation}
Here $\alpha_A$ is a coupling constant, and the superscript \textit{ind} means \textit{induced}. That is, $D^{\text{ind}}$ is a geometric covariant derivative different from the physical $D_{\mu}$ and $\tilde{D}_{\mu}$.

$A^{\text{ind}}_{\mu}$ is correspondingly an induced gauge potential; it is the connection on the principal bundle and therefore carries
the topological information of the defects. $A^{\text{ind}}_{\mu}$ can be expressed by $\phi$ through \eqref{ParallCond}:
\begin{equation}\label{Amu}
A^{\text{ind}}_{\mu}=\frac{\alpha_A}{2\pi}\frac{1}{2i\phi^*\phi}
\left(\phi^*\partial_{\mu}\phi-\partial_{\mu}\phi^*\phi\right),
\hspace*{15mm} * \text{ --- c.c.}.
\end{equation}
Eq.\eqref{Amu} takes the same form as the velocity field in quantum mechanics, hence it reproduces the so-called London assumption of superconductivity \cite{Liu:2002JPCM}.  From $A_{\mu}^{\text{ind}}$ the induced gauge field strength can be defined:
\begin{equation}
F^{\text{ind}}_{\mu\nu}=\partial_{\mu}A^{\text{ind}}_{\nu}-\partial_{\nu}A^{\text{ind}}_{\mu}. \label{FmunuInd}
\end{equation}
$A_{\mu}^{\text{ind}}$ and $F^{\text{ind}}_{\mu\nu}$ are to be used in the next section to construct the topological Hopf mapping degree
to measure entanglements between knotted cosmic strings.

$\phi$ is a complex scalar field, $\phi=\phi^1+i\phi^2$, with $\phi^1,\phi^2 \in \mathbb{R}$. A two-dimensional unit vector $n^a$ can be defined from $\phi^{1,2}$:
\begin{equation}
n^a=\frac{\phi^a}{\parallel \phi \parallel},\hspace*{15mm}a=1,2;~\left\| \phi \right\|^2=\phi^a\phi^a=\phi^*\phi.
\end{equation}
In terms of $n^a$ the potential and the field strength are given by
\begin{equation}
A^{\text{ind}}_\mu=\frac{\alpha_A}{2\pi}\epsilon_{ab}n^a\partial_\mu n^b,\hspace{10mm}
\hspace*{10mm} F^{\text{ind}}_{\mu\nu}=\frac{\alpha_A}{2\pi}2\epsilon_{ab}\partial_\mu n^a\partial_\nu n^b.
\end{equation}
Introducing a topological tensor current
\begin{equation}\label{Lgr}
j^{\mu\nu}=\frac{1}{2}\frac{1}{\sqrt{-g}}\epsilon^{\mu\nu\lambda\rho}F^{\text{ind}}_{\lambda\rho},
\end{equation}
one can prove \cite{Liu:2002} that $j^{\mu\nu}$ can be expressed by a $\delta$-function,
\begin{equation} \label{jmudelta}
j^{\mu\nu}=\frac{\alpha_A}{\sqrt{-g}}\delta^2(\mathbf{\phi})
D^{\mu\nu}\left(\frac{\phi}{x}\right),
\end{equation}
where the Jacobian $\epsilon^{ab}D^{\mu\nu}\left(\frac{\phi}{x}\right) =
\epsilon^{\mu\nu\lambda\rho}\partial_{\lambda}\phi^a\partial_{\rho}\phi^b$. In the deduction of \eqref{jmudelta} two formulae apply:
\begin{equation}
\partial_{\mu}n^a=\frac{\partial_{\mu}\phi^a}{\parallel\phi \parallel}+\phi^a \partial_{\mu}\frac{1}{\parallel\phi \parallel},\hspace*{15mm}\partial_a\partial_a\ln \left\| \phi \right\|=2\pi\delta^2(\mathbf{\phi}).
\end{equation}

In terms of the current $j^{\mu\nu}$ a geometric action is defined,
\begin{equation} \label{NamGotoActn1}
S=\int_{\mathbf{U}^4}L \sqrt{-g}d^4x, \hspace*{10mm} \text{with} \hspace*{10mm} L=\frac{1}{\alpha_A}\sqrt{\frac{1}{2}g_{\mu\lambda}g_{\nu\rho}j^{\mu\nu}j^{\lambda\rho}},
\end{equation}
$\mathbf{U}^4$ denoting the base manifold as a Riemann-Cartan space-time. $g$ is the determinant of the metric on $\mathbf{U}^4$. It will be shown shortly that $S$ is actually a Nambu-Goto action \cite{Nielsen:1973}.
Substituting \eqref{jmudelta} into \eqref{NamGotoActn1} we obtain an important form for the Lagrangian:
\begin{equation} \label{NambuGotoActn2}
L=\frac{1}{\sqrt{-g}}\delta^2(\mathbf{\phi}).
\end{equation}
Noticing the evaluation of the $\delta$-function
\begin{equation}
\delta^2(\mathbf{\phi}) \left\{
             \begin{array}{lll}
              =0, & & \text{iff }\phi^a\neq 0; \\
              \neq 0, & & \text{iff }\phi^a= 0,
             \end{array}
        \right. \hspace*{15mm} a=1,2,
\end{equation}
which implies that the desired topological non-triviality occurs only at the zeroes of the $\phi$ field, we need to solve the following zero-point equations to locate the topological defects:
\begin{equation} \label{phi12eqn}
\phi^1(x^1,x^2,x^3,x^0)=0,\hspace*{15mm} \phi^2(x^1,x^2,x^3,x^0)=0.
\end{equation}

According to the implicit function theorem, under a so-called regular condition $D^{\mu\nu}\left(\frac{\phi}{x}\right)  \neq 0$, the coupled equations in \eqref{phi12eqn} have a family of $N$ isolated solutions,
\begin{equation} \label{stringsolns}
 P_k:~~x^\mu=x^\mu_k(u^I),\hspace*{15mm} k=1,2,\cdots,N;~I=1,2.
\end{equation}
Each $P_k$, serving as a two-dimensional singular submanifold, is a world-sheet swept by a one-dimensional line defect. $u^1,u^2$ act as the intrinsic parameters of $P_k$, i.e., the spatial parameter $s$ and the temporal parameter $t$, respectively.

The Lagrangian $L$ then reads
\begin{equation} \label{Lagstrings}
L=\frac{1}{\sqrt{-g}}\sum_{k=1}^N W_k \int_{P_k}
\delta^4\left(x^{\mu} - x^{\mu}_k(s,t)\right) \sqrt{g_k} d^2u,
\end{equation}
where $g_k = \det\left(g_{IJ}\right)$, $g_{IJ}$ being the intrinsic metric of the submanifold $P_k \left(u^I\right)$. The $W_k$ is the winding number of the string $L_k$, $W_k=\beta_k \eta_k$, where $\beta_k$ is the Hopf index and $\eta_k = \pm 1$ the Brouwer degree. Thus, the action becomes a sum
\begin{equation} \label{actnstrings}
S=\sum_{k=1}^N W_k S_k \hspace*{10mm} \text{with}
\hspace*{10mm} S_k=\int_{P_k}\sqrt{g_k} d^2u,
\end{equation}
every $S_k$ corresponding to the area of $P_k$. Therefore $S_k$ is the Lagrangian of the $k$th world sheet, and \eqref{NamGotoActn1} virtually defines a Nambu-Goto action.

The motion of a string $L_k$ is investigated via the spatial component of the current $j^{\mu\nu}$,
\begin{equation} \label{topcurrspatial}
j^i=j^{0i}=\frac{\alpha_A}{\sqrt{-g}}\sum_{k=1}^{N}W_k\int_{l_k}
\frac{dx^i}{ds}\delta^3 \left(\mathbf{x}-\mathbf{x}_k (s)\right) ds, \hspace*{10mm} i= 1,2,3.
\end{equation}
As $\frac{dx^i}{ds}$ gives the velocity of the $k$th cosmic string,
\begin{equation} \label{velocitystring}
\frac{dx^i}{ds} = \frac{D^i \left(\frac{\mathbf{\phi}}{x}\right)}{D \left(\frac{\mathbf{\phi}}{u}\right)},
\end{equation}
the string evolution is governed by the following equation of motion:
\begin{equation} \label{stringevolutioneqn}
\frac{1}{\sqrt{g_k}}\frac{\partial}{\partial u^I}\left(\sqrt{g_k} g^{IJ} \frac{\partial x^{\mu}}{\partial u^J} \right)
+ g^{IJ} \Gamma^{\mu}_{\nu \lambda} \frac{\partial x^{\nu}}{\partial u^I} \frac{\partial x^{\lambda}}{\partial u^J} =0,
\hspace*{10mm} \mu,\nu=1,2,3,0;~I,J=1,2.
\end{equation}

\begin{figure}[H]%width=0.30\textwidth
      \centering
      \includegraphics[scale=0.23]{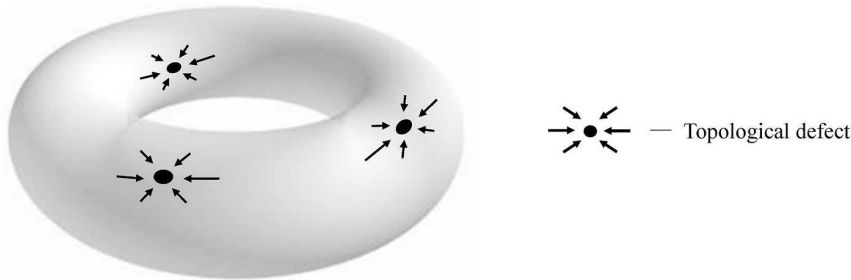}
      \caption{Consider for instance a base manifold $T^2 = S^1 \times S^1$, a torus, on which there exist three topological defects. A $\theta$-term is able to express only the topology of $T^2$ but not the defects; the topological nontriviality of the defects is reflected by the distribution of the $\phi$ field as a solution to the equation of motion.}
      \label{TorusDef}
  \end{figure}

In the end of this section a remark should be noted on the above theoretical analysis, eqs.\eqref{1stActn}--\eqref{FmunuInd}.
Usually the topological properties of a physical model is investigated by adding an extra term into a Lagrangian:
\begin{equation}
S = \int_{\mathcal{M}} \sqrt{-g} d^4x \left[ L_{\text{dyn}} + L_{\text{top}}  \right],
\end{equation}
where $L_{\text{dyn}}$ is the physical dynamical term, and $L_{\text{top}}$ a Lagrangian of topological invariant, such as a $\theta$-term containing the second Chern class. However, we argue that such a $\theta$-term is assigned to convey only the global topological information of the original base manifold $\mathcal{M}$ itself, without taking into consideration the new generated nontrivial topological defects existing whereon. For example, we can consider a base manifold, torus $T^2$, as in Fig.\ref{TorusDef}. A dynamical equation of motion together with a boundary condition gives rise to a number of topological defects.

\noindent  To study these defects we need to appeal to eqs.\eqref{1stActn}--\eqref{FmunuInd}, to solve out the $\phi$ field distribution from the EOM which delivers the topological information of the defects. Then a gauge potential $A^{\text{ind}}$ and a field tensor $F^{\text{ind}}$ are induced from $\phi$ through the parallel field condition, eq.\eqref{ParallCond}, $D^{\text{ind}}\phi =0$. \textit{This procedure follows up the routine of Riemannian geometry, where the Christoffel connection $\Gamma^{\mu}_{\nu \lambda}$ is achieved from the parallel field requirement,
$\nabla_{\mu} g_{\nu\lambda}= \partial_{\mu} g_{\nu\lambda} - \Gamma^{\rho}_{\mu \nu} g_{\rho\lambda}
- \Gamma^{\rho}_{\mu \lambda} g_{\nu\rho} =0$, and the Riemannian curvature $R^{\mu}_{\nu \lambda \rho}$ is then defined from $\Gamma^{\mu}_{\nu \lambda}$}. See the gauge potential decomposition theories of Faddeev, Cho, et al \cite{Faddeev:1999,Cho:1999jn}.

%%%%%%%%%%%%%%%%%%%%%%%%%%%%%%%%%
%%%%%%%%%%%%%%%%%%%%%%%%%%%%%%%%%
%%%%%%%%%%%%%%%%%%%%%%%%%%%%%%%%%

\section{Topological invariant of knotted cosmic strings}

Oriented knot-like topological excitations have been discovered in many areas such as nematic fields, magnetic flux tubes and semi-flexible polymers \cite{Martinez:2014,Forbes:2000,Marenz:2016}. Rich numerical simulation has been performed for string-net works of strings \cite{Vilenkin:1994,Moore:2001px}. Usually a highly tangled structure is unstable and intends to break up into less knotted sub-structures through a series of reconnection events, in which a topological complexity decreasing procedure is accompanied by energy release \cite{Kleckner:2015}. It is crucial to develop a powerful topological approach to detect knottedness of cosmic strings for the purpose of obtaining a better understanding of topology-energy relationship.

\subsection{Hopf mapping degree for knotted cosmic strings}

Our starting point is the topological Hopf mapping degree. In 1997 Faddeev and Niemi suggested a classical action of 3-dimensional Yang-Mills gauge theory to generate stable soliton solutions of finite energy and finite lengths \cite{Niemi:1997}. A soliton is a closed (or compactificated at infinity) string whose tangledness is characterized by a topological charge defined in a spatial volume $\Omega$:
\begin{equation} \label{eq:QCSdefn}
Q=\left(\frac{2\pi}{\alpha_A}\right)^2\frac{1}{4\pi}\int_\Omega \epsilon^{ijk}A^{\text{ind}}_iF^{\text{ind}}_{jk}d^3x, \hspace*{15mm} \epsilon^{ijk}=\epsilon^{0ijk},
\end{equation}
where $A^{\text{ind}}_i$ and $F^{\text{ind}}_{jk}$ are the spatial component of the above $A^{\text{ind}}_{\mu}$ and $F^{\text{ind}}_{\mu\nu}$, respectively. This charge $Q$ is a topological invariant of the Hopf mapping, $\pi_3(S^2)=\mathbf{Z}$.

This invariant may find applications in wide physical areas including particle physics, superfluidity and so on. From the prospective of fluid mechanics it corresponds to a key topological invariant, helicity: $H = \int_\Omega \mathbf{u} \cdot \mathbf{\omega} d^3x $, where $\mathbf{u}$ is the fluid velocity and $\mathbf{\omega}$ the vorticity, $\mathbf{\omega} = \nabla \times \mathbf{u}$.
Moffat and Ricca proved \cite{Moffatt:1992} that if the fluid contains a link $\mathcal{L}$ composed of $N$ component knots (i.e., vortex filaments), $\mathcal{L} = \{\gamma_k,~k=,1,\cdots,N\}$, then $H$ can be delivered by an algebraic sum of the (self-)linking numbers of the components $\gamma_k$'s,
\begin{equation}\label{CScharge2}
Q=2\pi\left[\sum_{k=1}^N W^2_k SL(\gamma_k)+2\sum_{k,l=1(k< l)}^N W_kW_l Lk(\gamma_k,\gamma_l) \right],
\end{equation}
where $W_k$ is the flux (i.e., topological charge) of the  $k$th filament $\gamma_k$, $SL(\gamma_k)$ the self-linking number of $\gamma_k$, and $Lk(\gamma_k,\gamma_l)$ the mutual linking number between $\gamma_k$ and $\gamma_l$. In \cite{Bekenstein:1992} Bekenstein transplanted this result into the study of intercommutation processes of cosmic strings; a strict verification of \eqref{CScharge2} in the context of cosmology was given in \cite{Liu:2002}.

\subsection{Weakness of the linking number}

\begin{figure}[H]
\centering
\subfigure[]{\includegraphics[width=3cm]{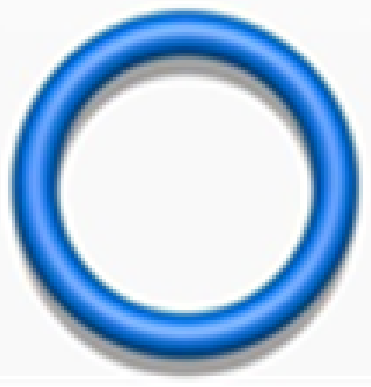}}
\hspace*{20mm}
\subfigure[]{\includegraphics[width=3cm]{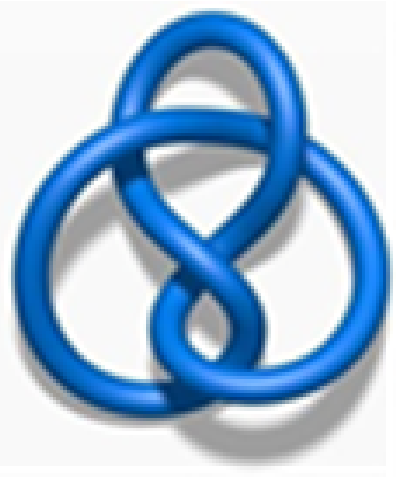}}
\hspace{5ex}
\caption{Different topological configurations share the same self-linking number $0$: (a) a circle (topologically trivial); (b) a figure-8 knot (topologically non-trivial).}
\label{Example1}
\end{figure}

\begin{figure}[H]\centering
\subfigure[]{\includegraphics[width=3cm]{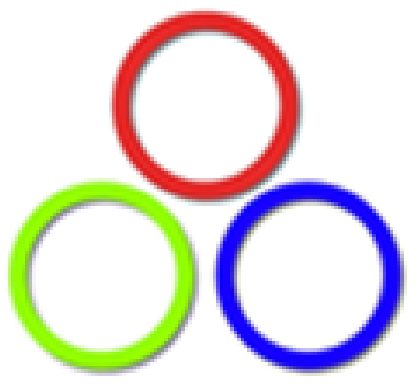}}
\hspace*{10mm}
\subfigure[]{\includegraphics[width=3cm]{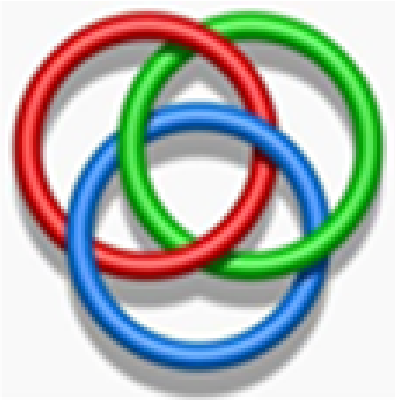}}
\hspace*{10mm}
\subfigure[]{\includegraphics[width=3cm]{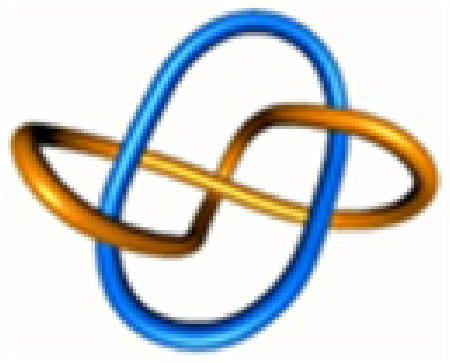}}
\caption{Different topological configurations share the same linking number $0$: (a) three disjoint trivial circles; (b) the Borromean rings; (c) the Whitehead links.}
\label{Example2}
\end{figure}

However, in knot theory (self-)linking numbers are actually a weak tool in characterizing knot topology. For example, as shown in Fig.\ref{Example1}, the self-linking number $0$ fails to distinguish a trivial circle and a figure-8 knot. In Fig.\ref{Example2}, the three disjoint circles in (a), the Borromean rings in (b) and the Whitehead links in (c) share the same mutual linking number $0$.

Therefore, in order to effectively characterize and distinguish topology of cosmic strings we need to develop new and stronger topological invariants. Our starting point is still the Hopf mapping degree $Q$, because helicity is the most important topological invariant in fluid mechanics, while the Hopf degree $Q$ has essential significance in geometry and topology. Indeed, the Hopf degree $Q$ of \eqref{eq:QCSdefn} can be re-written in the language of exterior differential as
\begin{equation}
Q = \frac{2\pi}{\alpha_A^2} \int_{\Omega} A^{\text{ind}} \wedge dA^{\text{ind}},
\qquad \text{with} \quad F^{\text{ind}} = dA^{\text{ind}},
\end{equation}
which is exactly an Abelian Chern-Simons action. It is well known that the Chern-Simons topological quantum field theory provides a field theoretical framework for knot theory \cite{Witten:1989} from which rich knot invariants arise, such as the Jones, Kauffman, HOMFLYPT and other knot polynomials, as well as the Vassiliev finite-type invariants, etc. \cite{Birman:1993,Lin:2000,Barnatan:1995}. Hence the Hopf degree $Q$ is promising in revealing more topological information of cosmic strings other than linking numbers.

\section{The HOMFLYPT polynomial}

In this section the HOMFLYPT knot polynomial invariant will be constructed in terms of the Hopf mapping degree $Q$. Our emphasis will be placed on the fact that the two parameters of HOMFLYPT, $z$ and $a$ (see \eqref{HOM-Relation-2} later), are obtained from the writhes and twists of the cosmic strings, respectively. This agrees to the common knowledge of fluid mechanics that writhing and twisting have equal contributions to helicity \cite{Ricca:1992,Moffatt:1992}.

In the light of \eqref{Lgr} and \eqref{topcurrspatial}, which imply considering zero-width cosmic strings, the Hopf mapping degree of \eqref{eq:QCSdefn} becomes
\begin{equation}
Q = \frac{2\pi}{\alpha^2_A}\int_\Omega A^{\text{ind}}_i j^{i} \sqrt{-g} d^3x = \frac{1}{\alpha_A} \sum_k W_k \oint _{\gamma_k} A_i^{\text{ind}} dx^i .
\end{equation}
Since the strings $\gamma_k$'s form a link $\mathcal{L} = \{ \left. \gamma_k \right| k=1,\cdots,N \}$, this $Q$ provides a measurement for the entangledness of the strings. In the following, for simplicity, let us rescale the dimensional coefficient $\alpha_A =1$, and set the topological charge of each string $W_A =1$. Thus,
\begin{equation}
Q = \sum_k \oint _{\gamma_k} A_i^{\text{ind}} dx^i = \oint _{\mathcal{L}} A_i^{\text{ind}} dx^i,
\hspace*{15mm} Q \text{ --- dimensionless}.
\end{equation}

For the purpose of revealing deeper essence of the Hopf degree other than linking numbers, we propose to consider the exponential form \cite{Liu:2012}
\begin{equation}\label{topoinexponent}
  e^{Q\left(\mathcal{L}\right)}= \exp\left(\oint_{\mathcal{L}}  A^{\text{ind}}_{i} dx^{i} \right).
\end{equation}
The advantages of this exponential form $e^{Q\left(\mathcal{L}\right)}$ in comparison to the non-exponential $Q\left(\mathcal{L}\right)$ are the following:
\begin{itemize}
  \item \textsf{Additivity of integration strands and factorization of exponential form}:

  A link $\mathcal{L}$ is able to be divided into the sum of a number of strands $S_1,S_2,S_3,\cdots$ in the configuration space, i.e., $\mathcal{L} = S_1 \oplus S_2 \oplus S_3 \oplus \cdots$, as shown in Fig.\ref{Strandecompn}(a).
  The line integral over $\mathcal{L}$ is thus additive:
  \begin{equation}\label{Strandecompn-1}
    \oint_{\mathcal{L}} = \int_{S_1} + \int_{S_2} + \int_{S_3} + \cdots .
  \end{equation}
  The addition in the power leads to factorization of the exponential, $
    e^{\oint_{\mathcal{L}}} = e^{\int_{S_1}}e^{\int_{S_2}}e^{\int_{S_3}}\cdots
  $, which provides us a way to construct the formal parameter $a$ in the skein relations of the Kauffman bracket polynomial.
  \begin{figure}[H]
      \centering
      \includegraphics[width=0.60\textwidth]{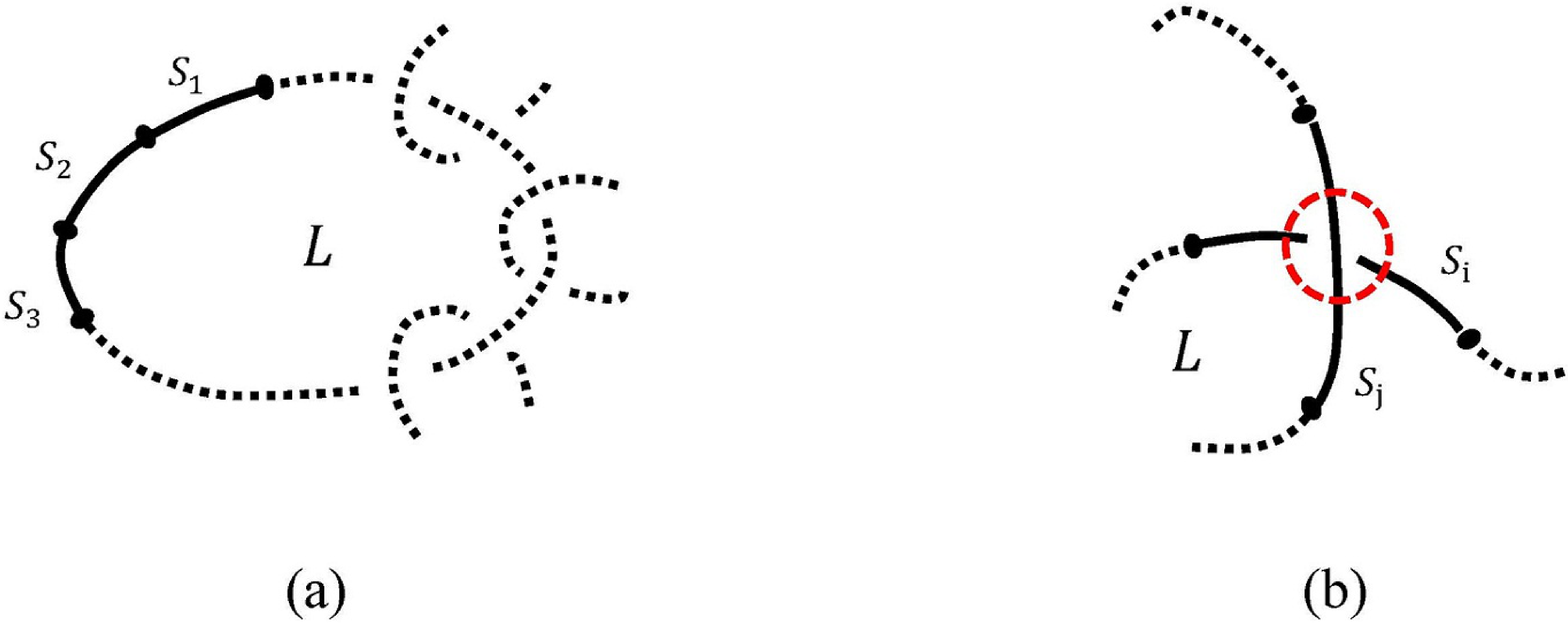}
      \caption{(a) A link $\mathcal{L}$ is additive, i.e., it can be treated as the sum of a number of strands $S_1,S_2,S_3,\cdots$. (b) The interaction between two strands $S_i$ and $S_j$ can be described by their correlations, with the dashed red-circled region as the interaction area \cite{Liu:Preparation}.}
      \label{Strandecompn}
  \end{figure}

  \item \textsf{Interaction between strings described by correlation of strands}:

  Examining the Taylor-expansion of $e^{Q\left(\mathcal{L}\right)}$, we have
   \begin{eqnarray}
       && e^{Q \left( \mathcal{L} \right)} = e^{\oint_{\mathcal{L}}} = e^{\int_{S_1} + \int_{S_2} + \int_{S_3}\cdots} \notag \\
       &=&1+ Q \left( \mathcal{L} \right) + \frac{1}{2!}\sum_{i,j}\int_{S_i} \int_{S_j} + \frac{1}{3!}\sum_{i,j,k}\int_{S_i} \int_{S_j} \int_{S_k} + \cdots. \label{TaylorexpLineIntgl}
   \end{eqnarray}
   Obviously, the first term reproduces $Q\left(\mathcal{L}\right)$, while the higher terms provide a mechanism to study (self-)interactions of cosmic strings, as shown in Fig.\ref{Strandecompn}(b) \cite{Liu:Preparation}.

\end{itemize}

\subsection{Derivation of  R-polynomial}

Our strategy of proof in this section below is first to construct the so-called Kauffman R-polynomial \cite{Kauffman:1987}, and then derive the HOMFLYPT knot polynomial.

The R-polynomial for oriented knots is introduced by L.H. Kauffman \cite{Kauffman:1987}:
\begin{gather}
R\left(\PICorientcircleright\right)=1,      \label{R-poly-1}\\
  R\left(\PICorientopengammaplus\right)=a  R\left(\PICorientline\right),  \hspace{15mm} R\left(\PICorientopengammaminus\right)=a^{-1}  R\left(\PICorientline\right),     \label{R-poly-2}\\
  R\left(\PICorientpluscross\right)-R\left(\PICorientminuscross\right)=
  zR\left(\PICorientLRsplit\right),   \label{R-poly-3}
\end{gather}
where $a$ and $z$ are two independent parameters, and \PICorientpluscross , \PICorientminuscross ~ and \PICorientLRsplit ~ are three almost-the-same links which differe only at one crossing site: over-, under- and non-crossings, respectively.

Eqs.\eqref{R-poly-1} to \eqref{R-poly-3} serve as a recurrence program for computing the R-polynomial of a knot/link. \eqref{R-poly-1} is the starting point of recurrence; in the following we will construct the skein relations \eqref{R-poly-3} and \eqref{R-poly-2} subcessively in two subsections, to construct the two independent parameters $z$ and $a$ by analyzing the contributions of writhing and twisting, respectively.

%%%%%%%%%%%%%%%%%%%%%%%%%%%%%%
%%%%%%%%%%%%%%%%%%%%%%%%%%%%%%
%%%%%%%%%%%%%%%%%%%%%%%%%%%%%%

\subsubsection{Parameter $z$ arising from writhing contribution}

In order to derive \eqref{R-poly-3}, which is concerned about oriented knots, let us turn to un-oriented knots/links first; after achieving for the un-oriented the Kauffman bracket polynomial, we will come back to the oriented.

\begin{figure}[H]
\centering
\includegraphics[width=0.82\textwidth]{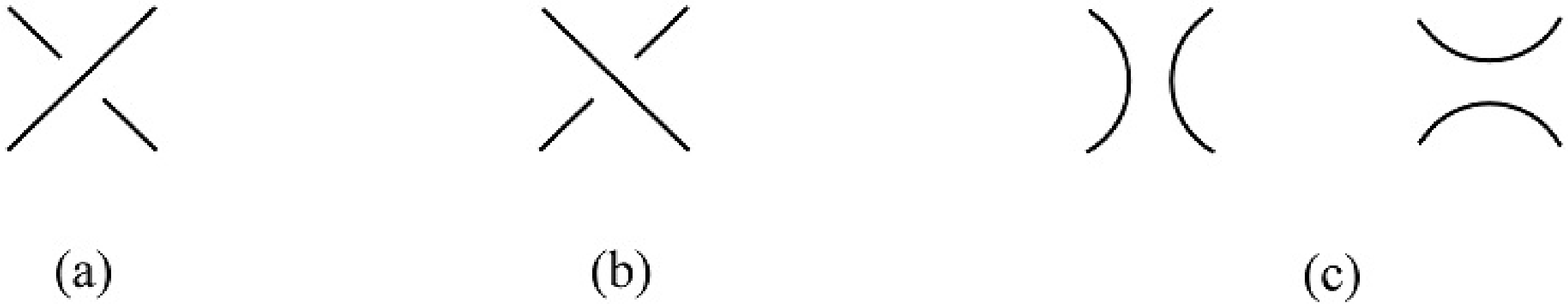}
\caption{The different states of four almost-the-same knots/links which differ only at one particular crossing site: (a) over-crossing denoted as $L_+$; (b) under-crossing denoted as $L_-$; (c) non-crossings: (\textit{left}) $L_0$, left-right open;  (\textit{right}) $L_{\infty}$, up-down open.}
\label{crossingstates}
\end{figure}

For un-oriented knots/links, let us start from the four basic states at a crossing site: the over-crossing, under-crossings, and the left-right-open and up-down-open non-crossings, as shown in Fig.\ref{crossingstates}.

\noindent Moreover, let us introduce two important writhing loop notations for convenient use in later text:
\begin{equation}
\gamma_+ = \PICclosegammaplus,\hspace*{15mm}\gamma_- = \PICclosegammaminus.
\end{equation}

In order to find the relationship between crossings and non-crossings, we need to perform a tehnique named ``adding/subtracting imaginary local paths''. Without loss of generality, let us consider for example the $L_+$, \PICunorientpluscross :
\begin{itemize}
\item If \PICunorientpluscross is endowed with proper orientations to form an over-crossing \PICorientpluscross , it can be related to \PICorientLRsplit via
\begin{figure}[H]
\centering
\includegraphics[width=0.50\textwidth]{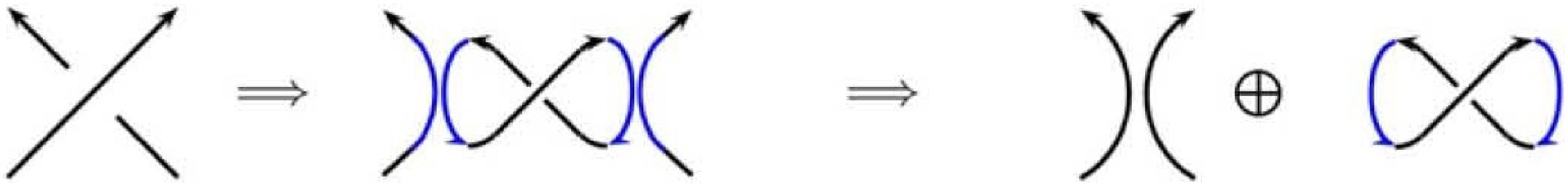}
\end{figure}
where the colored strands are imaginarily added local paths which cancel each other.

\item If endowed with orientations the other way, to form an under-crossing \PICorientminuscrossrotated , it can be related to \PICorientUDsplit via
\begin{figure}[H]
\centering
\includegraphics[width=0.50\textwidth]{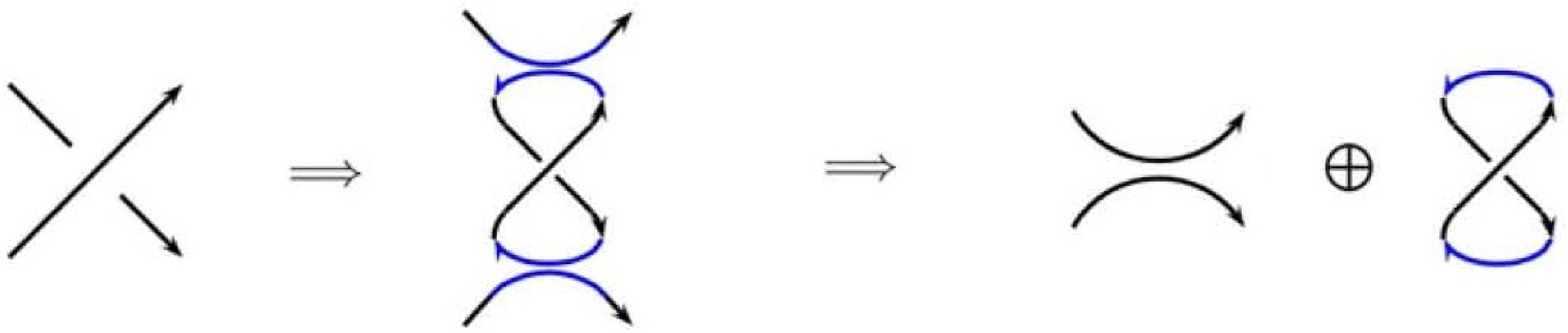}
\end{figure}

\end{itemize}

\noindent That means, for the unoriented $L_+$ , we have two ways of splitting:
\begin{figure}[ht]
\centering
\includegraphics[width=3in]{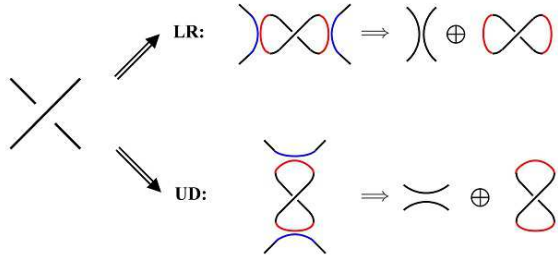}
\caption{Two channels to decompose the crossing state $L_+$: Left-right(LR) and up-down(UD), through adding a pair of local parallel strands denoted by red and blue colors. \textit{Upper row}: LR splitting, with $L_+$ split into a non-crossing $L_0$ and a writhing loop $\gamma_+$. \textit{Lower row}: UD splitting, with $L_+$ split into a non-crossing $L_\infty$ and a writhing loop $\gamma_-$ }
\label{CrossingDecom}
\end{figure}

\noindent Mathematically, \PICunorientpluscross means
\begin{equation}
\left\langle \PICunorientpluscross \right\rangle = \left\langle L_+ \right\rangle = e^{\oint_{L_+}},
\hspace*{10mm} \text{where the notation } e^{\oint_{\mathcal{L}}} = \exp \left( \oint_{\mathcal{L}} A_i^{\text{ind}} dx^i  \right) \text{is used for short.}
\end{equation}
Thus, if using a so-called ergodic statistical hypothesis --- the two splittings of Figure \ref{CrossingDecom} play an equal role in the expansion of $\left\langle L_+ \right\rangle$ --- we have
\begin{equation}
\left\langle L_+ \right\rangle = \left\langle L_+ \right\rangle_{\text{LR}}
+ \left\langle L_+\right\rangle_{\text{UD}} , \label{D0}
\end{equation}
where
\begin{equation}
\left\langle L_+ \right\rangle_{\text{LR}} = e^{\oint_{L_0 \oplus \gamma_+}}
= e^{\oint_{L_0}}e^{\oint_{\gamma_+}},
\hspace*{15mm}
\left\langle L_+ \right\rangle_{\text{UD}} = e^{\oint_{L_{\infty} \oplus \gamma_-}}
= e^{\oint_{L_{\infty}}}e^{\oint_{\gamma_-}}.
\label{D0-12}
\end{equation}
Denoting
\begin{eqnarray}
&& \left\langle \PICunorientLRsplit \right\rangle = e^{\oint_{L_0}},
 \hspace*{24mm}
 \left\langle \PICunorientUDsplit \right\rangle = e^{\oint_{L_{\infty}}},
 \label{D0-3}\\
&& \alpha = e^{\oint_{\gamma_+}} = e^{\oint_{\PICclosegammaplus}} , \hspace*{15mm}
\alpha^{-1} = e^{\oint_{\gamma_-}} = e^{\oint_{\PICclosegammaminus}} = e^{-\oint_{\PICclosegammaplus}},
\label{D0-4}
\end{eqnarray}
\eqref{D0} becomes
\begin{equation}\label{D1}
 \left\langle \PICunorientpluscross \right\rangle =
 \alpha \left\langle \PICunorientLRsplit \right\rangle +
 \alpha^{-1} \left\langle \PICunorientUDsplit \right\rangle.
\end{equation}
Similarly, $L_-$ has a decomposition
\begin{equation}\label{D2}
 \left\langle \PICunorientminuscross \right\rangle =
 a^{-1} \left\langle \PICunorientLRsplit \right\rangle +
 a \left\langle \PICunorientUDsplit \right\rangle, \hspace*{10mm}
 \text{with~~~} \left\langle \PICunorientminuscross \right\rangle = \left\langle L_- \right\rangle = e^{\oint_{L_-}}.
\end{equation}

Now we are at the stage of going back to the oriented case. Endowing $L_{\pm}$ in \eqref{D1} and \eqref{D2} with orientations, we have
\begin{eqnarray}
 \left\langle \PICorientpluscross \right\rangle =
    \alpha \left\langle \PICorientLRsplit \right\rangle +
    \alpha^{-1} \left\langle \PICorientUDinconsistentsplit \right\rangle, \label{OrienKuaf1}\\
     \left\langle \PICorientminuscross \right\rangle =
    \alpha^{-1} \left\langle \PICorientUDinconsistentsplit \right\rangle +
    \alpha \left\langle \PICorientLRsplit \right\rangle. \label{OrienKuaf12}
    \end{eqnarray}
$\left\langle \PICorientUDinconsistentsplit \right\rangle$ is an inconsistent term and can be safely cancelled by combining \eqref{OrienKuaf1} and \eqref{OrienKuaf12}:
\begin{equation}
  \alpha \left\langle \PICorientpluscross \right\rangle -\alpha      \left\langle \PICorientminuscross \right\rangle=\left(\alpha^2-\alpha^{-2}\right) \left\langle \PICorientLRsplit \right\rangle.
  \label{OrienKuafmid}
\end{equation}
\eqref{OrienKuafmid} immediately leads to the desired third skein relation of the R-polynomial, \eqref{R-poly-3}. Indeed, in terms of the standard position of \cite{Kauffman:2001}, the R-polynomial of a knot diagram $\mathcal{L}$ is readily obtained from the bracket polynomial through compensating a writhe of $\mathcal{L}$, $w=w\left[\mathcal{L}\right]$:
\begin{equation}\label{RKrelation}
  R\left[\mathcal{L}\right]=\alpha^w \left<\mathcal{L}\right>.
\end{equation}
Applying \eqref{RKrelation} to non-crossing, over-crossing and under-crossing states, we have
\begin{eqnarray}
R\left( \PICorientLRsplit\right)&=&\alpha^{w}\left\langle\PICorientLRsplit \right\rangle, \label{Rpoly1}\\
R\left(\PICorientpluscross\right)&=&\alpha^{w+1}\left\langle\PICorientpluscross \right\rangle,\hspace{1cm} \label{Rpoly2}\\
R\left(\PICorientminuscross\right)&=&\alpha^{w-1}\left\langle\PICorientminuscross \right\rangle,\label{Rpoly3}
\end{eqnarray}
then \eqref{OrienKuafmid} is at once rewritten as
\begin{equation}
R\left(\PICorientpluscross\right)-R\left(\PICorientminuscross\right) =
 \left(\alpha^2-\alpha^{-2}\right)R\left(\PICorientLRsplit\right). \label{OrienRpolymid}
\end{equation}
Defining $k=\alpha^2$ and $z=k-k^{-1}$, \eqref{OrienRpolymid} steadily presents the third skein relation of the R-polynomial, \eqref{R-poly-3},
\begin{equation*}
  R\left(\PICorientpluscross\right)-R\left(\PICorientminuscross\right)=
  zR\left(\PICorientLRsplit\right). \label{Rwrithe}
\end{equation*}
We stress here that the parameter $\alpha$ and consequently the $z$ arise from the writhing, $\alpha = e^{\oint_{\gamma+}}$ and $z = e^{2\oint_{\gamma_+}} - e^{2\oint_{\gamma_-}}$.

\subsubsection{Parameter $a$ arising from twist contribution}

\begin{figure}[H]
\centering
\includegraphics[width=0.5\textwidth]{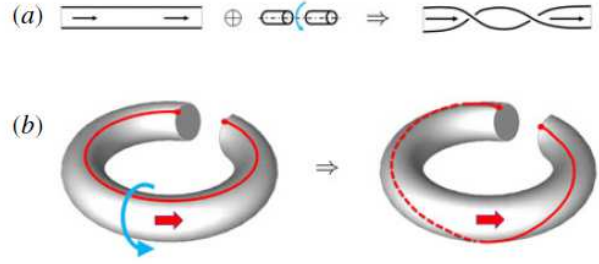}
\caption{(a) A cosmic string with intrinsic twist can be regarded as  a twisted ribbon, which is mathematically identical to an untwisted ribbon plus a twisting operation. (b) In the study of physical knots, a practical approach of treating physical twists is the so-called Dehn surgery; here a $2\pi$-twist of a torus is shown.}
\label{FigDehn1}
\end{figure}

\begin{figure}[H]
\centering
 \includegraphics[width=0.5\textwidth]{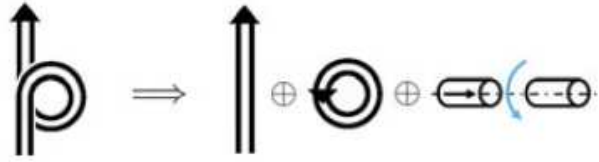}
 \caption{Consider a twisted ribbon (\textit{on the left}). With the aid of Figure \ref{FigDehn1}(b), it can be decomposed into the sum of an untwisted ribbon, a trivial annulus and a positive twisting operation (\textit{on the right}).}
      \label{FigDehn2}
\end{figure}

The second skein relation of the R-polynomial, \eqref{R-poly-2}, is abel to be derived from the computation of twist contribution. The intrinsic twist of a strand of a cosmic string is shown in Figure \ref{FigDehn1}.
In the context of physical knots, the Dehn surgery is considered as an idealized technique to treat insertion/deletion of intrinsic twists of knots \cite{Adams:1999}, where twisting operation is found to contribute an independent twist helicity \cite{Moffatt:1990}. In our present study of cosmic strings, towards \eqref{R-poly-2} which describes the difference between $R\left(\PICorientopengammaplus\right)$ and $R\left(\PICorientline\right)$, we need to consider \PICorientopengammaplus as a twisted ribbon as shown in Figure \ref{FigDehn2}.

\noindent As happened in fluid mechanics, the twisted ribbon is identical to the sum of an untwisted ribbon, a trivial annulus and a positive twisting operation, which is mathematically expressed as
\begin{equation}
\exp\left(\oint_{\PICorientopengammaplus}\right) =
\exp\left(\oint_{\PICorientline}\right)
\exp\left(\oint_{\PICcircle}\right)
\exp\left(\oint_{\text{positive twisting operation}}\right),
\end{equation}
where $\exp\left(\oint_{\PICcircle}\right) = 1$ by definition. Noticing $\exp\left(\oint_{\text{positive twisting operation}}\right)$ corresponds to a Dehn surgery that leads to an independent twist-contribution to the Hopf degree $Q\left( \mathcal{L} \right)$, we need to introduce a new, independent parameter as
\begin{equation}
a = \exp\left(\oint_{\text{positive twisting operation}}\right).
\end{equation}
Thus, in the language of R-polynomials, we arrive at
\begin{equation}
 R\left(\PICorientopengammaplus\right) = a  R\left(\PICorientline\right),\hspace*{15mm}
 R\left(\PICorientopengammaminus\right) = a^{-1}  R\left(\PICorientline\right), \label{RTwist}
\end{equation}
where $ R\left(\PICorientopengammaplus\right) = \exp\left(\oint_{\PICorientopengammaplus}\right)$,
$ R\left(\PICorientopengammaminus\right) = \exp\left(\oint_{\PICorientopengammaminus}\right)$,
$R\left(\PICorientline\right) = \exp\left(\oint_{\PICorientline}\right)$, and $a^{-1}$ is the negative twisting operation,
\begin{equation}
a^{-1} = \exp\left(\oint_{\text{negative twisting operation}}\right) = \exp\left(- \oint_{\text{positive twisting operation}}\right).
\end{equation}
The two formulae in \eqref{RTwist} steadily present the second skein relation of the R-polynomial, \eqref{R-poly-2}. We stress that the parameter $a$ stems from the twisting operation, totally independent from the above writhing parameter $z$.

%%%%%%%%%%%%%%%%%%%%%%%%%%%%%%%
%%%%%%%%%%%%%%%%%%%%%%%%%%%%%%%
%%%%%%%%%%%%%%%%%%%%%%%%%%%%%%%

\subsection{Derivation of the HOMFLYPT polynomial from the R-polynomial}

The R-polynomial obtained above is regular-isotopic, i.e., sensitive to a type-I Reidemeister move, demonstrated by the second skein relation \eqref{R-poly-2}. In contrast, the desired HOMFLYPT polynomial by definition should be ambient-isotopic, i.e., insensitive to a type-I Reidermester move,
\begin{equation}
  P\left(\PICorientopengammaplus\right)=  P\left(\PICorientline\right),\hspace{15mm}
  P\left(\PICorientopengammaminus\right)=P\left(\PICorientline\right). \label{HOM-mid}
\end{equation}
Thus, in the standard position below to derive the HOMFLYPT polynomial from the R-polynomial \cite{Kauffman:2001}, a compensation factor should be introduced:
\begin{equation}\label{HOM-relation}
  P\left(a,z\right)=a^{-w}R\left(a,z\right),
\end{equation}
where $a^{-w}$ is a compensation factor:
\begin{equation}
  w\left[\PICorientopengammaplus\right] = +1, \hspace*{10mm}
  w\left[\PICorientopengammaminus\right]=-1;  \hspace*{10mm}
  w\left[\PICorientcircleright\right] = w\left[\PICorientline\right]=w\left[\PICorientLRsplit\right]=0.
\end{equation}
Then, $P\left( \mathcal{L} \right)$ becomes the desired polynomial satisfying the following skein relations of HOMFLYPT:
\begin{eqnarray}
&& P\left(\PICorientcircleright\right)=1, \label{HOM-Relation-1} \\
&& aP\left(\PICorientpluscross\right)-a^{-1}P\left(\PICorientminuscross\right)=zP\left(\PICorientLRsplit\right),
\label{HOM-Relation-2}
\end{eqnarray}
where the two parameters $z$ and $a$ are independent parameters arising from the contributions of writhing and twisting of cosmic strings, respectively.

\section{Degeneration of HOMFLYPT to the Jones and Alexander-Conway polynomials}
\subsection{Derivation of the Jones polynomial}

The HOMFLYPT polynomial is a generalized version of the Jones polynomial, which degenerates to the Jones when
\begin{equation}
a=\tau^{-1}, \hspace*{15mm} z=\tau^{-\frac{1}{2}}-\tau^{\frac{1}{2}}. \label{evaluatnHOMJones}
\end{equation}
The skein relations of the Jones polynomial read
\begin{eqnarray}
&& V \left( \PICcircle \right) = 1, \\
&& \tau^{-1} V\left(\PICorientpluscross\right) - \tau V\left(\PICorientminuscross\right)
= (\tau^{-\frac{1}{2}}-\tau^{\frac{1}{2}}) V\left(\PICorientLRsplit\right). \label{heliRe11}
\end{eqnarray}
It can be verified that $V\left( \mathcal{L} \right)$ is an ambient-isotopic invariant, i.e., insensitive to a type-I Reidermester move,
\begin{equation}
V\left(\PICorientopengammaplus\right)=V\left(\PICorientopengammaminus\right)=V\left(\PICorientline\right). \label{heliRe10}
\end{equation}

The evaluation \eqref{evaluatnHOMJones} implies $z$ and $a$ are no longer independent parameters but satisfy a special relationship:
\begin{equation}
z^2 = a + a^{-1} -2.
\end{equation}
Hence the Jones polynomial is a single-parameter polynomial.

\begin{figure}[H]
\centering
\includegraphics[width=0.5\textwidth]{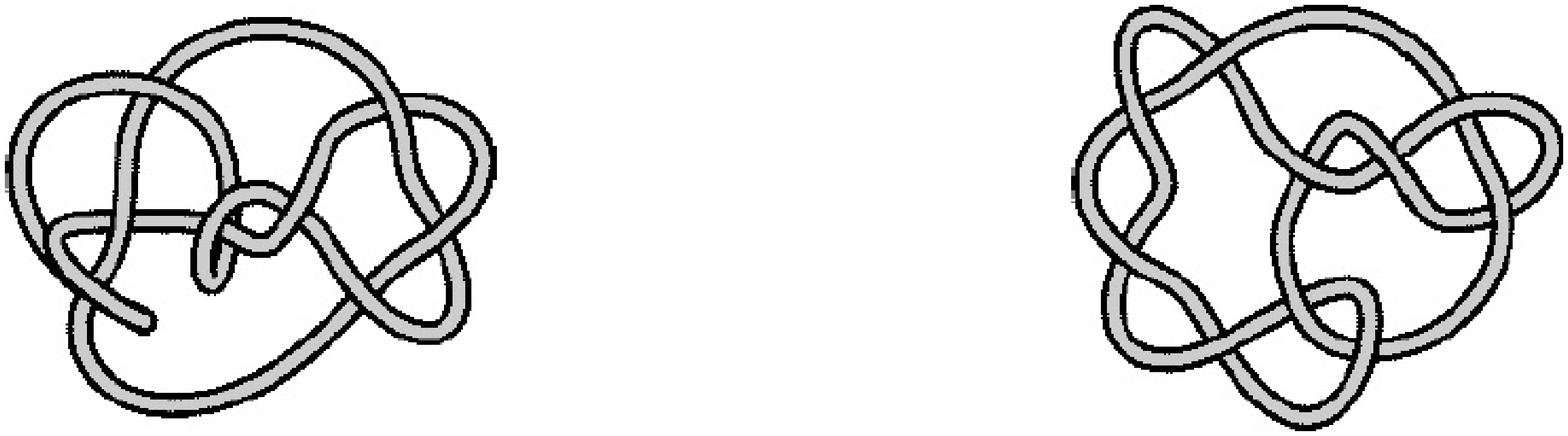}
\caption{Two knots which share the same Jones polynomial but have different HOMFLYPT polynomials. {\it Left}: $10$-$022$; {\it right}: $10$-$035$.}
\label{Jonescnttell}
\end{figure}

As a knot topological invariant, the HOMFLYPT polynomial is stronger than the Jones in the power of distinguishing knots and links. For example, the Jones polynomial fails to distinguish the two knots $10$-$022$ and $10$-$035$  in Figure \ref{Jonescnttell}, but the HOMFLYPT does.

\subsection{Derivation of the Alexander-Conwey polynomial}

If evaluating the parameters $z$ and $a$ in another way:
\begin{equation}
a=1, \hspace*{15mm} z = t^{\frac{1}{2}} - t^{-\frac{1}{2}},
\end{equation}
the HOMFLYPT polynomial degenerates to the so-called Alexander-Conway polynomial, which satisfies the following skein relations
\begin{eqnarray}
&& \Delta \left( \PICcircle \right) = 1, \\
&& \Delta\left(\PICorientpluscross\right) - \Delta\left(\PICorientminuscross\right)
= \left(t^{-\frac{1}{2}} - t^{\frac{1}{2}} \right) \Delta\left(\PICorientLRsplit\right). \label{Alex-ConwPoly}
\end{eqnarray}
The Alexander-Conway is weaker than the Jones polynomial in distinguishing knots and links.

%%%%%%%%%%%%%%%%%%%%%%
%%%%%%%%%%%%%%%%%%%%%%
%%%%%%%%%%%%%%%%%%%%%%

\section{Examples}

The HOMFLYPT and Jones polynomials of some typical knots and links are listed as follows.

\begin{center}
\begin{tabular}{|c|l|l|}
\hline
& HOMFLY polynomial & Jones polynomial \\ \hline
$\overset{n\text{ circles}}{\overbrace{\PICcircle \cdots \PICcircle }}$ & $\delta ^{n-1}$, where $\delta =\frac{a-a^{-1}}{z}$ & $\delta = -\tau ^{%
\frac{1}{2}}-\tau ^{-\frac{1}{2}}$ \\ \hline
$\mathcal{L}\sqcup \PICcircle $ & $
P\left( \mathcal{L}\right) \delta $ \hspace*{10mm} for any oriented link $\mathcal{L}$ & $V\left( \mathcal{L}\right) \delta $
\\ \hline
$\mathbf{H}_{+}$ & $a^{-1}z+\left( a^{-1}-a^{-3}\right) z^{-1}$ & $%
-\tau ^{\frac{1}{2}}-\tau ^{\frac{5}{2}}$ \\ \hline
$\mathbf{H}_{-}$ & $-az-\left( a-a^{3}\right) z^{-1}$ & $-\tau ^{-%
\frac{1}{2}}-\tau ^{-\frac{5}{2}}$ \\ \hline
$\mathbf{T}^{L}$ & $2a^{2}+a^{2}z^{2}-a^{4}$ & $-\tau
^{-4}+\tau ^{-3}+\tau ^{-1}$ \\ \hline
$\mathbf{T}^{R}$ & $2a^{-2}+a^{-2}z^{2}-a^{-4}$ & $-\tau
^{4}+\tau ^{3}+\tau $ \\ \hline
$\mathbf{F}^{8}$ & $a^{-2}-1-z^{2}+a^{2}$ & $\tau ^{-2}-\tau
^{-1}+1-\tau +\tau ^{2}$ \\ \hline
$\mathbf{W}$ & $%
-a^{-1}z^{-1}-a^{-1}z+az^{-1}+2az+az^{3}-a^{3}z$ & $\tau ^{-\frac{7}{2}%
}-2\tau ^{-\frac{5}{2}}+\tau ^{-\frac{3}{2}}-2\tau ^{-\frac{1}{2}}+\tau ^{%
\frac{1}{2}}-\tau ^{\frac{3}{2}}$ \\ \hline
$\tilde{\mathbf{W}}$ & $%
az^{-1}+az-a^{-1}z^{-1}-2a^{-1}z-a^{-1}z^{3}+a^{-3}z$ & $\tau ^{\frac{7}{2}%
}-2\tau ^{\frac{5}{2}}+\tau ^{\frac{3}{2}}-2\tau ^{\frac{1}{2}}+\tau ^{-%
\frac{1}{2}}-\tau ^{-\frac{3}{2}}$ \\ \hline
$\mathbf{B}$ & $%
a^{-2}z^{-2}-a^{-2}z^{2}+a^{2}z^{-2}-a^{2}z^{2}-2z^{-2}+2z^{2}+z^{4}$ & $%
-\tau ^{-3}+3\tau ^{-2}-2\tau ^{-1}+4-2\tau +3\tau ^{2}-\tau ^{3}$ \\ \hline
\end{tabular}
\end{center}

\noindent where

\begin{center}

\begin{tabular}{lllllll}
$\mathbf{H}_+$ &---& Hopf link, & \hspace*{15mm} & $\mathbf{H}_-$ &---& Hopf link (different orientations);  \\
$\mathbf{T}^L$ &---& Trefoil knot (left handed),  & & $\mathbf{T}^R$ &---& Trefoil knot (right handed); \\
$\mathbf{F}^8$ &---& Figure-of-eight knot; & & &&\\
$\mathbf{W}$ &---& Whitehead link, & & $\tilde{\mathbf{W}}$ &---& Whitehead link (mirror image);\\
$\mathbf{B}$ &---& Borromean rings. &  & &&
\end{tabular}
\end{center}

%%%%%%%%%%%%%%%%%%%%%%
%%%%%%%%%%%%%%%%%%%%%%
%%%%%%%%%%%%%%%%%%%%%%

\section{Conclusion}

In this paper closed superconducting cosmic strings (SCSs) are studied as oriented knotted line defects. Our emphasis is placed on the Hopf topological mapping degree of the strings. This invariant is an Abelian Chern-Simons type action, from which the HOMFLYPT knot polynomial has been constructed. We show that the two independent parameters of HOMFLYPT, $z$ and $a$, correspond to the writhing and twisting contributions, respectively. This new method is topologically stronger than the traditional (self-) linking number method which fails to detect essential topology of knots sometimes, opening a new door to the study of physical reconnections of SCSs as a complex system. Finally, the degenerations of the HOMFLYPT to the Jones and Alexander-Conway polynomials are presented, and the knot polynomials of some typical knot and link configurations are listed for example.

It is remarked that in this paper SCSs are studied as classical mathematical curves with zero-width. In practice, the thickness of a real SCS should be taken into consideration, especially in the study of string intercommunications where energy and entropy changes count \cite{Achucarro:2010ub}.

%%%%%%%%%%%%%%%%%%%%%%
%%%%%%%%%%%%%%%%%%%%%%
%%%%%%%%%%%%%%%%%%%%%%

\section{Acknowledgements}
We are indebted to Professor Renzo L. Ricca for helpful discussions. This work was financially supported by the National Natural Science Foundation of China (NSFC, No.11572005), and the Beijing Overseas Talent Aggregation Project Fellowship (Haiju Project, Youth).

\end{document}